\documentclass[prl,twocolumn,aps,groupedaddress,showpacs]{revtex4}

\usepackage{amsmath,amsfonts,amssymb}
\usepackage{graphicx}
\usepackage{bbm,bm}

\newcommand{\be}{\begin{eqnarray}}
\newcommand{\ee}{\end{eqnarray}}
\newcommand{\eins}{\mbox{$1 \hspace{-1.0mm}  {\bf l}$}}

\begin{document}

\title{Ground state approximation for strongly interacting systems in arbitrary dimension}

\author{S.\ Anders,$^1$ M.~B.\ Plenio,$^{2,3}$ W.\ D\"ur,$^{1,4}$ F.\ Verstraete,$^5$ and H.-J.\ Briegel$^{1,4}$}

\affiliation{$^1$ Institut f{\"u}r Theoretische Physik, Universit{\"a}t Innsbruck,
Technikerstra{\ss}e 25, A-6020 Innsbruck, Austria\\
$^2$ QOLS, Blackett Laboratory, Imperial College London, Prince Consort Road, London SW7 2BW, UK\\
$^3$ Institute for Mathematical Sciences, Imperial College London, 53 Exhibition Road, London SW7 2BW, UK\\
$^4$ Institut f\"ur Quantenoptik und Quanteninformation der \"Osterreichischen Akademie der Wissenschaften, Innsbruck, Austria\\
$^5$ Institute for Quantum Information, California Institute of Technology, Pasadena, CA 91125, USA}

\date{February 27, 2006}

\begin{abstract}
We introduce a variational method for the approximation of ground
states of strongly interacting spin systems in arbitrary
geometries and spatial dimensions. The approach is based on
weighted graph states and superpositions thereof. These states
allow for the efficient computation of all local observables (e.\,g.
energy) and include states with diverging correlation length and
unbounded multi-particle entanglement. As a demonstration we apply
our approach to the Ising model on 1D, 2D and 3D square-lattices.
We also present generalizations to higher spins and continuous-variable 
systems, which allows for the investigation of lattice
field theories.
\end{abstract}


\maketitle


Strongly correlated quantum systems are of central interest in
several areas of physics. Exotic materials such as high-$T_c$
superconductors and quantum magnets exhibit their remarkable
properties due to strong quantum correlations, and experimental
breakthroughs with e.g. atomic gases in optical lattices
provide a perfect playground for probing strongly correlated
quantum systems. The main obstacle in understanding the behavior
of those quantum systems is the difficulty in simulating
the effective Hamiltonians that describe their properties.
In most cases, the strong correlations in the exponentially large Hilbert space render
an exact solution infeasible, and attacking the problem by
numerical means requires sophisticated techniques such as
quantum Monte Carlo (QMC) methods or the density matrix renormalization group (DMRG) approach \cite{Wh91,Sc04}.

QMC methods suffer from the sign problem which makes them
inappropriate for the description of fermionic and frustrated quantum
systems. DMRG is a variational approach that provides approximations to ground states, thermal states and dynamics of many--body systems.
Recent insight from entanglement theory have lead to an improved
understanding of both the success and the limitations of this
approach. Indeed, the accuracy of the method is closely linked to
the amount of entanglement in the approximated states
\cite{Vi03,Ve04}. Matrix product states \cite{FNW92}, which provide the
structure underlying DMRG, are essentially one--dimensional and
the entanglement entropy of these states is limited by the dimension $D$
of the matrices, which in turn is directly linked to the
computational cost \cite{Vi03,Sc04}. Hence a successful
treatment of systems with bounded entanglement, e.g. one--dimensional,
non--critical spin systems with short range
interactions, is possible, while the method is inefficient for
systems with an unbounded amount of entanglement, e.g. critical
systems and systems in two or more dimensions. Promising
generalizations that can deal with higher dimensional
systems have been reported recently \cite{Ve04b,Vi05}.
However, the computational effort and complexity increases with the dimension of the system.
In addition, the amount of block-wise entanglement of the states used in Ref.\ \cite{Ve04b} still scales proportional
at most to the surface of a block of spins, whereas in general a scaling in proportion
to the volume of the block is possible. Such a scaling can in fact
be observed for disordered systems \cite{Ca05} or systems with
long--range interactions \cite{Du05}.

Here we introduce a new variational method using states with
intrinsic long-range entanglement and no bias towards a geometry
to overcome these limitations. We first illustrate our
methods for spin-1/2 systems, and then generalize them to
arbitrary spins and infinite dimensional systems such as
harmonic oscillators. In finite dimensions, the method is
based on a certain class of multiparticle--entangled spin states,
\textit{weighted graph states} (WGS) and superpositions thereof. WGS are a
$O(N^2)$ parameter family of $N$--spin states with the following
properties: (i) they form an (overcomplete) basis, i.e. any
state can be represented as a superpositions of WGS; (ii) one can
efficiently calculate expectation values of any localized
observable, including energy, for any WGS; (iii) they correspond
to weighted graphs which are independent of the geometry and hence
adaptable to arbitrary geometries and spatial dimensions; (iv) the
amount of entanglement contained in WGS may be arbitrarily high,
in the sense that the entanglement between any block of $N_A$
particles and the remaining system may be $O(N_A)$ and the
correlation length may diverge.

Note that (iii) and (iv) are key properties in which this approach
differs from DMRG and its generalizations and which suggest a potential
for enhanced performance at least in certain situations, while
(ii) is necessary to efficiently perform
variations over this family. In the following we will outline how
we use superpositions of a small number of WGS as variational
ansatz states to find approximations to ground states of strongly
interacting spin systems in arbitrary spatial dimension.



{\em Properties of WGS.}
WGS are defined as states of $N$ spin-$1/2$ (or qubits),
that result from applying phase gates 
$U_{ab}(\varphi_{ab}) =
\operatorname{diag}(1,1,1,e^{-i\varphi_{ab}})$ onto each pair of
qubits $a,b\in\{1,2,\dots,N\}$  of a tensor product of $\sigma_x$-eigenstates
$|+\rangle =(|0\rangle + |1\rangle)/\sqrt{2}$, followed by a
single-qubit filtering operation $D_a =
\operatorname{diag}(1,e^{d_a})$, $d_a\in\mathbb{C}$ and a general
unitary operation $U_a$
\begin{equation}
    |\Psi_{\Gamma,\mathbf{d},U}\rangle \propto \prod_{a=1}^N U_a D_a
    \prod_{b=a+1}^N U_{ab}(\varphi_{ab}) |+\rangle^{\otimes N}.
\end{equation}
The phases $\varphi_{ab}$ can be associated with a weighted graph
with a real symmetric adjacency matrix $\Gamma_{ab} =
\varphi_{ab}$. For convenience, we define a deformation vector
$\mathbf{d}=(d_1,d_2,\ldots ,d_N)$ and $U \equiv{\bigotimes}_a
U_a$. 
%
The deformations make WGS as used in this letter slightly
more general than the WGS used in Refs.\ \cite{Ca05,Du05} where
$d_a=0$. One can conveniently rewrite
$|\Psi_{\Gamma,\mathbf{d},U}\rangle$ as
\begin{equation}
\label{WGS_simple}
    |\Psi_{\Gamma,\mathbf{d},U}\rangle \propto  U \sum_{\bm s}
    e^{-i{\bm s}^T \Gamma {\bm s}/2 + \mathbf{d}^T {\bm s}} |{\bm s}\rangle,
\end{equation}
where the sum runs over all computational basis states, which are labelled
with the binary vector ${\bm s}=(s_1,s_2, \ldots ,s_N)^T$. Our class of variation states comprises superpositions of WGS of the form
\begin{equation}
    \label{basis} |\Psi\rangle \propto \sum_{i=1}^m
    \alpha_i|\Psi_{\Gamma,\mathbf{d}^{(i)},U}\rangle,
\end{equation}
i.~e. the superposed states differ only in their deformation vector $\mathbf{d}^{(i)}$,
while the adjacency matrix $\Gamma$ and the unitary $U$ are fixed.
Such a state is specified by $N(N-1)/2 + 3N +2(N+1)m  = O(N^2)$ real parameters.

We now proceed to verify the properties set out in the
introduction. For property (i), observe that for any fixed
$\Gamma$ and $U$, all possible combinations of $D_a \in
\{\sigma_z^{(a)},\eins^{(a)}\}$ lead to an orthonormal basis (note
that $\sigma_z^{(a)},\eins^{(a)}$ commute with $U_{ab}$). Hence
{\em any} state $|\Psi\rangle$ can be written in the form Eq.\ 
(\ref{basis}) for sufficiently large $m \leq 2^N$, which shows the
exhaustiveness of the description.

The relevance of employing deformations lies in the
observation that only $|\Psi\rangle$ of the form of Eq.\ (\ref{basis}) permit the
efficient evaluation of the expectation values of localized
observables $A$, i.e. satisfy property (ii). For
simplicity we restrict our attention to observables of the form
\begin{equation}
     A=\sum_{a<b} A_{ab} + \sum_a A_a, \label{Observable}
\end{equation}
where $A_{ab}$ has support
on the two spins $a,b$. The method can be easily adopted to any
observable that is a sum of terms with bounded support.
To compute ${\rm tr}(A |\Psi\rangle\langle\Psi|)= \sum_{a<b} {\rm
tr}(A_{ab}\rho_{ab}) + \sum_a {\rm tr}(A_a \rho_a)$ it is
sufficient to determine the reduced density operators $\rho_{ab}$
and $\rho_a$.

For a single WGS ($m=1$)we obtain $\rho_{12}=(U_1\otimes U_2)(\sum r_{\bm s,\bm t} |{\bm s}\rangle
\langle {\bm t}|) (U_1\otimes U_2)^\dagger$ with
\begin{equation}
r_{\bm s,\bm t}=e^{-i \gamma}
    \prod_{c=3}^N
    \left(1+e^{d_c + d_c^* -i\sum_{e=1}^{2}(s_e-t_e)\Gamma_{ec}}\right)
    \label{reduced}
\end{equation}
and $\gamma= \sum_{a,b=1}^{2}\Gamma_{ab} (s_as_b - t_at_b)+
\sum_{a=1}^{2} (d_a s_a + d_a^* t_a)$.
This generalizes the formula for WGS without deformation obtained in Ref.\ \cite{Du05}.
Eq.\ (\ref{reduced}) demonstrates that for any WGS, the reduced
density operator of two (and one) spins can be calculated with a
number of operations that is linear in the system size $N$, as
opposed to an exponential cost for a general state.

A straight-forward generalization of Eq.\ (\ref{reduced}) allows one to calculate
two--qubit reduced density matrices for superpositions of the form of Eq.\ (\ref{basis})
in time $O(m^2 N)$. Therefore the expectation value of an observable $A$ of
the form of Eq.\ (\ref{Observable}) with $K$ terms requires $O(m^2 K N)$
steps. This implies that even for
Hamiltonians where all spins interact pairwise (and
randomly), i.e. $K = N(N-1)/2$, the expectation value of the energy for our ansatz
states can be obtained in $O(m^2 N^3)$ steps. For short--range
interaction Hamiltonians, this reduces to $O (m^2 N^2)$. The total
number of parameters (and memory cost) scales as $O(N^2 + m N)$,
which can be further reduced by employing symmetries.

Regarding (iii) and (iv), one can easily adopt a WGS to any given
geometry by a proper choice/restriction of the adjacency matrix
$\Gamma$. A state corresponding to a linear cluster state \cite{BrRa01}, for
instance, will have only $\Gamma_{a,a+1} \neq 0$, while
$\Gamma_{a,a+l} \neq 0$ would correspond to longer-ranged
correlations. Different choices of $\varphi_{ab}$ lead to very
different (entanglement) properties: For
$\varphi_{ab}=|\mathbf{x}_a-\mathbf{x}_b|^{-\beta}$, where $\mathbf{x}_a$ denotes the spatial
coordinates of spin $a$, one obtains diverging correlation length
for two-point correlations, while block--wise entanglement can
either be bounded or grow unboundedly, depending on the choice of
$\beta$ \cite{Du05}. Similarly, more complicated geometries on
lattices in higher spatial dimensions can be chosen.

{\em Variational method.} Any state of the form Eq.\ (\ref{basis})
with $m=\operatorname{poly}(N)$ permits the efficient calculation
of expectation values of any two--body Hamiltonian $H$. A good
approximation to the ground state is then obtained by numerical
optimization of the parameters characterizing the state. Starting
from random parameters, one descends to the nearest minimum using
a general local minimizer (we used L-BFGS \cite{By95}). Another
approach that we found to work well is to keep all parameters
fixed except for either those corresponding to (i) one local
unitary $U_a$, (ii) one phase gate $U_{ab}(\varphi_{ab})$ or (iii)
the deformation vector $d_a^{(j)}$ for one site $a$. In each case,
the energy as a function of this subset of parameters turns out to
be a quotient of quadratic forms, which can be optimized using the
generalized-eigenvalue (Rayleigh) method. A similar result holds
for the superposition coefficients $\alpha_j$. One then optimizes
with respect to these subsets of parameters in turns until
convergence is achieved. If one increases $m$ stepwise, one ---somewhat 
surprisingly--- does not get stuck in local minima.

A significant reduction of the number of parameters and the
computational costs may be achieved by exploiting symmetries, or
by adapting $\Gamma$ to reflect the geometrical situation. For
instance, for systems with short range interactions and finite
correlation length, one might restrict the range of the weighted
graph, i.e. $\Gamma_{ab}=0$ if $|\mathbf{x}_a-\mathbf{x}_b| \geq r_0$. This reduces
the number of parameters describing the WGS from $O(N^2)$ to
$O(N)$. For translationally invariant Hamiltonians, a better
scheme is to let $\Gamma_{ab}$ depend only on $|\mathbf{x}_a-\mathbf{x}_b|$. This
reduces the number of parameters to $O(N)$ as well, and it seems
to hardly affect the accuracy of the ground state approximation.
Hence, it allows one to reach high numbers of spins $N$ and thus to
study also 2D and 3D systems of significant size. Trading accuracy
for high speed one may even use a fully translation-invariant
ansatz, where also $D_a$ and $U_a$ are constant and independent of
$a$. In the latter case, for Hamiltonians with only
nearest-neighbor interactions, the expectation value of the energy
can be obtained by calculating only a {\em single} reduced density
operator, and the computational cost to treat 2D [and 3D] systems
of size $N=L^2$ [$N=L^3$] turns out to be of $O(L)$ rather than
$O(N)$.

\begin{figure}[ht]
\begin{minipage}{0.23\textwidth}
\includegraphics[height=3.2cm]{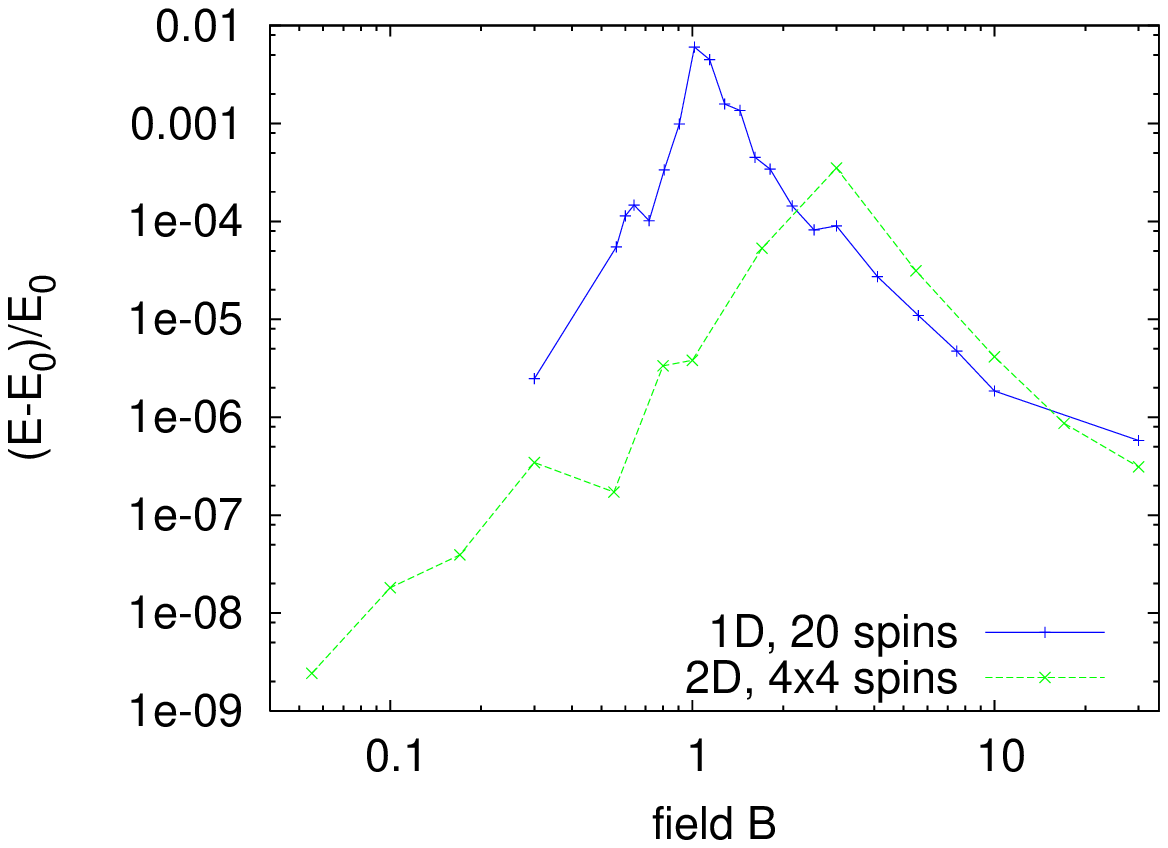}
\end{minipage}\hspace{2mm}%
\begin{minipage}{0.23\textwidth}
\includegraphics[height=2.9cm]{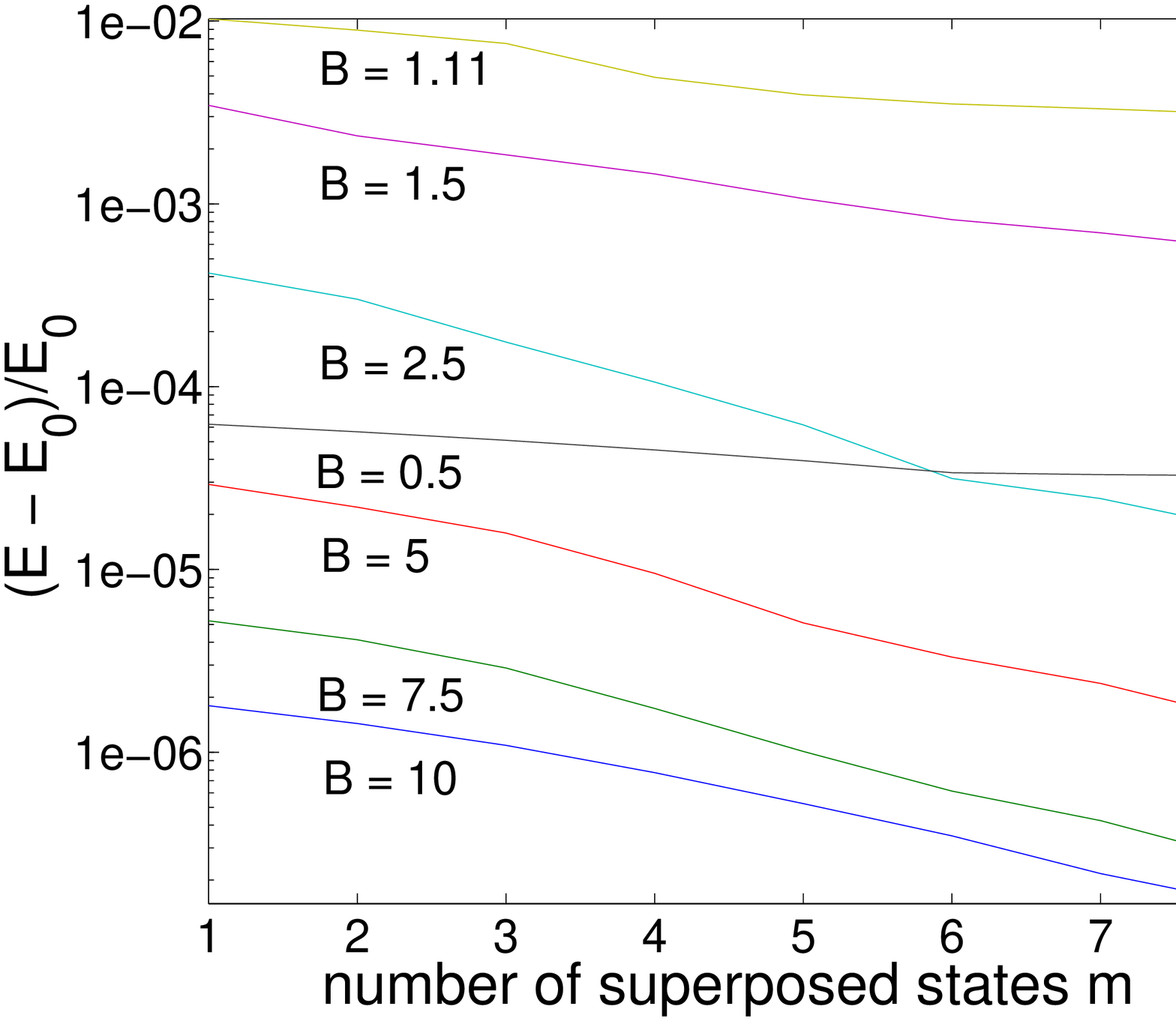}
\end{minipage}

\caption{\label{Exact}(Color online.) (a) Relative deviation from exact
ground state energy for Ising chain with $N=20$ (blue) and
$4\times 4$ 2D lattice (green) with periodic boundary conditions
as function of magnetic field $B$ (calculated using BFGS
minimization with symmetrized phases, $m \leq 6$). 
(b) 1D Ising chain with
$N=20$. Improvement of relative deviation from ground state energy
as function of number of superposed states $m$ for various field
values $B$ (calculated using Rayleigh minimization without
symmetrized phases). }
\end{figure}

\begin{figure}[hbt]
\includegraphics[height=3cm]{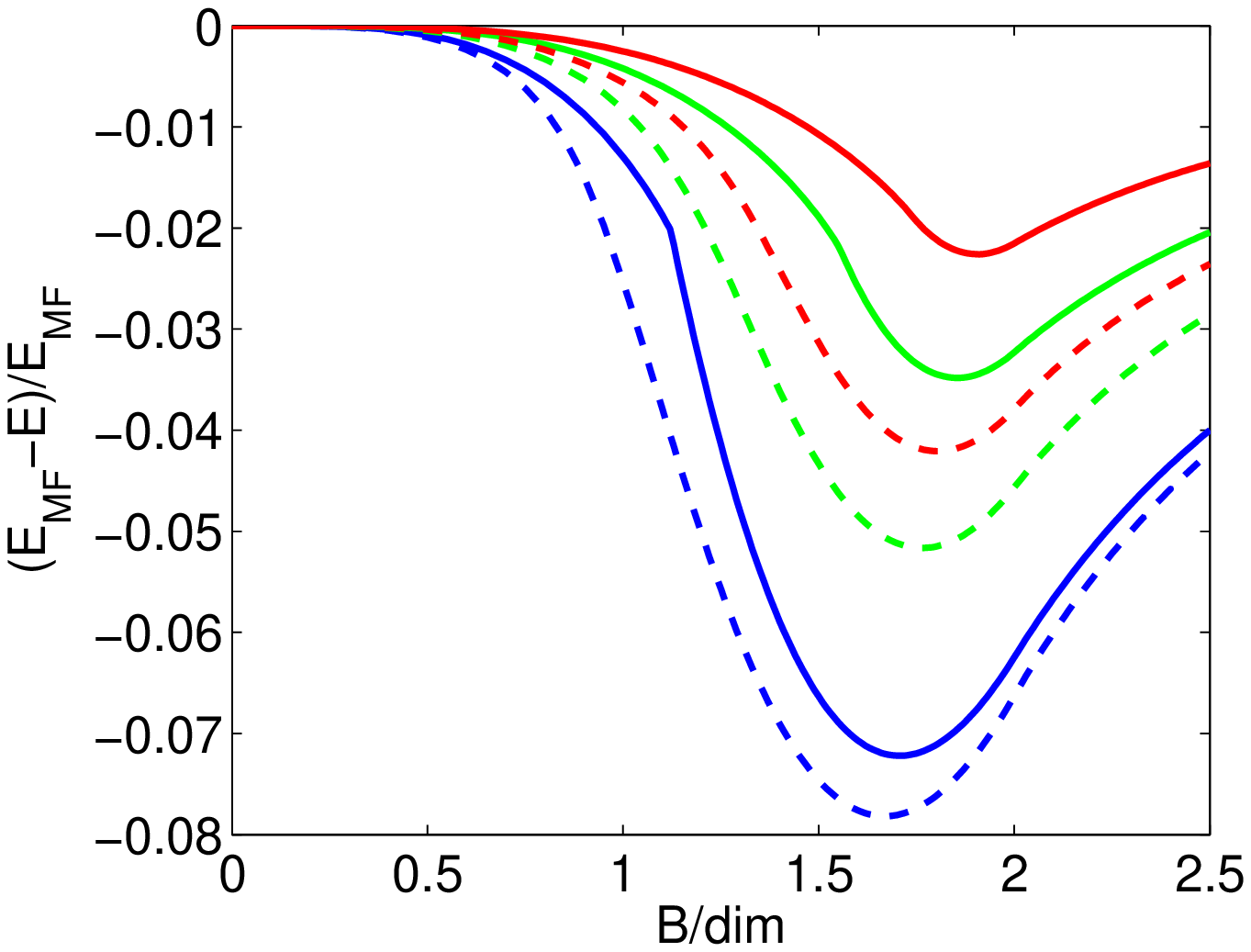}
\includegraphics[height=3.2cm]{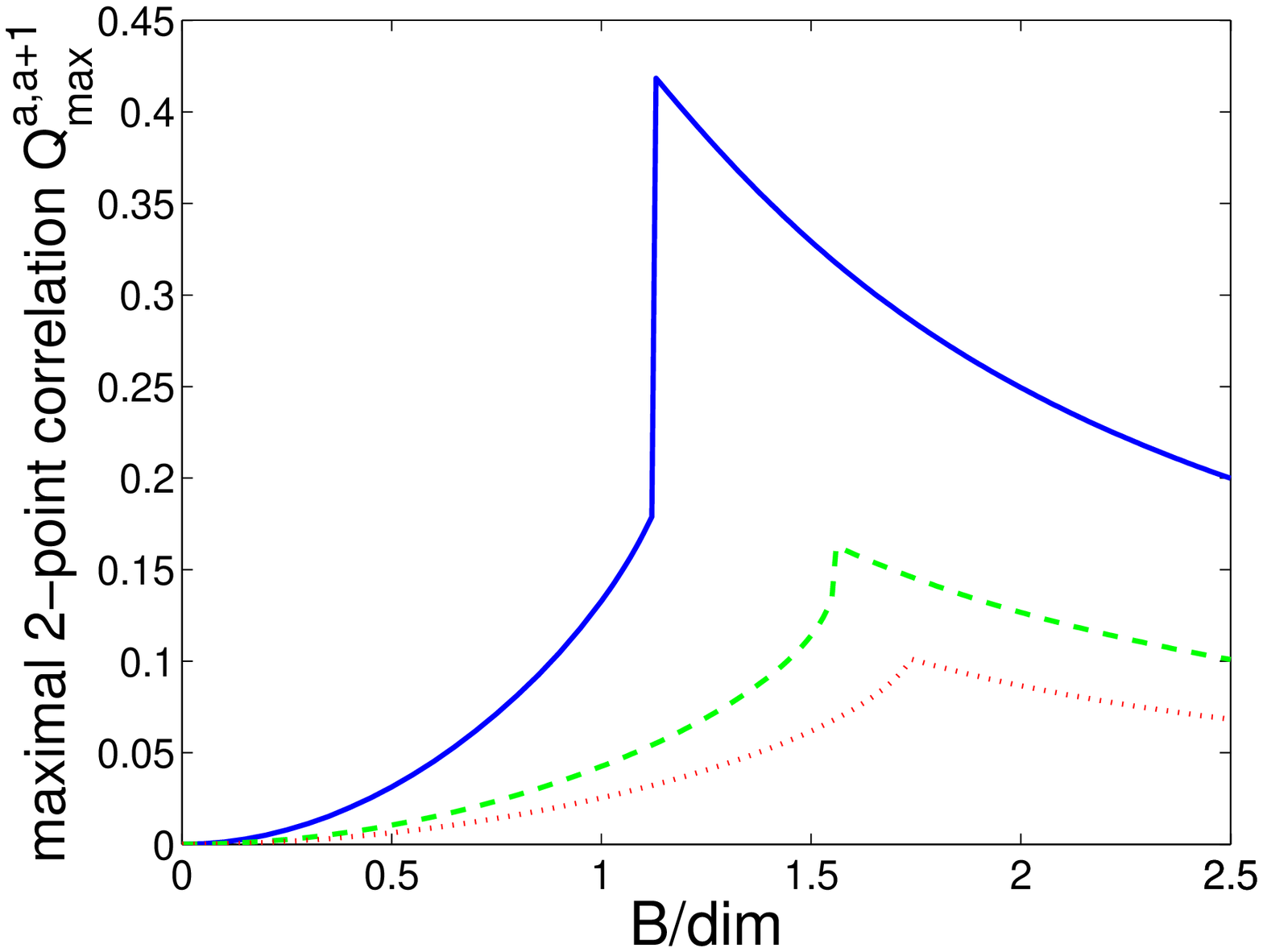}
\caption{\label{GroundState}(Color online.) Ising model in 1D
(blue) with $N=30$, 2D (green) with $N=30 \times 30 =900$ and 3D
(red) with $N=30 \times 30 \times 30 = 27000$ spins arranged as
chain, square, and cubic lattice, respectively,
for fully symmetric ansatz states with $\varphi_{ab}=f(|x_a-x_b|)$, $d_a=1$ as
function of magnetic field $B/{\rm dim}$, where ${\rm dim}$ is dimension of lattice.
(a) Relative deviation of ground state energy $(E_{MF}-E)/E_{MF}$ per
bond from to mean field approximation $E_{MF}$ (solid),
and of Anderson bound $(E_{MF}-E_A)/E_{MF}$ (dashed). Translational invariance is reduced
by using $U_1 \not= U_2$ (alternating). (b) maximal two-point
correlation $Q_{\rm max}^{a,a+1}$ for nearest neighbors. }
\label{Datasymmetric}
\end{figure}

{\em Demonstration. The Ising model.} Our method allows us to
determine, with only moderate computational cost, an upper bound
on the ground state energy of a strongly interacting system of
arbitrary geometry. Together with the Anderson lower bound, one
can hence obtain a quite narrow interval for the ground state
energy and observe qualitative features of the ground state
\cite{footnoteDMRG}. To illustrate our method, we have applied it
to the Ising model in 1D, 2D and 3D with periodic boundary
conditions, described by the Hamiltonian
\begin{equation}
H=-\sum_{\langle a,b\rangle} \sigma_z^{(a)}\sigma_z^{(b)} - B \sum_a \sigma_x^{(a)}
\end{equation}
where $\langle a,b\rangle$ denotes nearest neighbors. For a spin
chain with $N=20$, and a 2D lattice of size $4 \times 4$ we
compared our numerical ground state approximation with exact
results (Fig.\ \ref{Exact}a). We have also performed calculations for larger
2D systems up to $14 \times 14$. We note that the accuracy can be
further improved by increasing $m$ (see Fig.\ \ref{Exact}b). In fact
our numerical results suggest an exponential improvement with $m$.
We have also tested the fully translation invariant ansatz with
distance dependent phases, constant $d_a$ and
alternating $U_a$ for 1D, 2D and 3D systems of size $N=30$,
$N=900$ and $N=27000$ respectively (see Fig.\ \ref{Datasymmetric}).
There, for lack of a refernce value for the exact ground state, we compare with the
Anderson bound obtained by calculating the exact ground state
energy $E_A$ for system size $N=15,3^2, 2^3$ respectively. In
the 2D and especially the 3D case it is not expected that the
Anderson bound is particularly tight and may lead to a
significantly underestimation of the precisions achieved by our
approach. The states approximated with this simple ansatz
also show qualitatively essential features of the exact ground
state. As an example, the maximal two-point correlation function
$Q_{\rm max}^{a,a+1}$ (where the two point correlation functions
are defined as
$Q_{\alpha,\beta}^{a,b}=\langle\sigma_\alpha^{(a)}\sigma_\beta^{(b)}\rangle-
\langle\sigma_\alpha^{(a)}\rangle\langle\sigma_\beta^{(b)}\rangle$)
is plotted against the magnetic field $B$ in Fig.\ 
\ref{Datasymmetric}b. Strong indication for the occurrence of a
phase transition can be observed: the correlations significantly
increase around $B \approx 1.1,3.12,5.22$ in 1D, 2D, 3D
respectively. This is in good
agreement with estimates employing sophisticated power series
expansions for the {\em infinite} systems or Pad{\'e} approximants
based on large scale numerical simulations, which expect the
critical points at $B=1, 3.04, 5.14$ \cite{critpoint}. We also
remark that the approximated states show a scaling of block-wise
entanglement proportional to the surface of the block, i.\,e.
$S_{N_A} \approx \beta_B L^{{\rm dim}-1}$, where $\beta_B$ is some
constant depending on magnetic field $B$, $N_A=L^{\rm dim}$ and
${\rm dim}$ is the spatial dimension. We can estimate $\beta_B$
and find that it significantly increases near the critical point.

{\em Generalizations: } Our approach can be adapted directly to
spin-$\frac{n}{2}$ systems using the representation Eq.\ 
(\ref{WGS_simple}). There the sum over binary vectors $\mathbf{s}$ with $s_i=0,1$
has to be changed to $n$-ary vectors $\mathbf{s}$ with $s_i=0,1,...,n-1$ and
the corresponding matrices/vectors $\Gamma,\mathbf{d},U$ have to be
modified accordingly. However, the limit $n \to \infty$ to infinite
dimensional systems is both problematic and impractical, as the
computational effort increases with $n$. For continuous variable
systems we thus choose a closely related but slightly different
approach.

The description of field theories on lattices generally leads to
infinite-dimensional subsystems such as harmonic oscillators. A
Klein-Gordon field on a lattice for example possesses a
Hamiltonian quadratic in position and momentum operators $X$ and
$P$ whose ground state is Gaussian \cite{Plenio CDE 05}. This
suggests that techniques from the theory of Gaussian state
entanglement (see \cite{Eisert P 03} for more details) provide the
most natural setting for these problems. To this end, consider $N$
harmonic oscillators and the vector, $\mathbf{R}=
(R_1,...,R_{2N})^T=(X_1,P_1,...,X_{N},P_{N})^T.$ The canonical
commutation relations then take the form $[R_j,R_k]=i
\sigma_{jk}$ with the symplectic matrix $\sigma$. All
information contained in a quantum state $\rho$ can then be
expressed equivalently in terms of the characteristic function
$\chi_{\rho}(\bm{\xi}) = \operatorname{tr}[\rho W(\bm{\xi})]$ where
$\bm{\xi}\in{\mathbbm{R}}^{2N}$ and $W(\bm{\xi}) = \exp(i \bm{\xi}^{T} \sigma
\mathbf{R})$. Then, expectation values of polynomials of $X$ and $P$ can be
obtained as derivatives of $\chi$. For Gaussian states, i.e. states
whose characteristic function is a Gaussian $\chi_{\rho}(\bm{\xi}) =
\chi_{\rho}(0) e^{-\frac{1}{4}\bm{\xi}^T \gamma \bm{\xi} + \mathbf{D}^T \bm{\xi}}$, where
here $\gamma$ is a $2N\times 2N$-matrix and $\mathbf{D}\in{\mathbbm{R}}^{2N}$ is
a vector, these expectation values can be expressed efficiently as
polynomials in $\gamma$ and $\mathbf{D}$. On the level of wave functions a
pure Gaussian state is given by $|F,G;\mathbf{a}\rangle = C\int_{\mathbbm{R}^{N}}
d^Nx e^{-\frac{1}{2}\mathbf{x}^T(F-iG)\mathbf{x} + \mathbf{a}^T
\mathbf{x}} |\mathbf{x}\rangle$ where $F$ and $G$ are real symmetric matrices, $\mathbf{a}$ is a
vector, $C$ is the normalization and
\begin{eqnarray}
    \gamma = \left(\begin{array}{cc} F + GF^{-1}G  & -GF^{-1}\\
                                     -F^{-1}G     & F^{-1} \end{array}\right)
    ,\;    \mathbf{D} = \left(\begin{array}{c} GF^{-1}\mathbf{a} \\
    -F^{-1}\mathbf{a} \end{array}\right).
\end{eqnarray}
Now, we may consider coherent superpositions $|\psi\rangle=\sum_i
\alpha_i |G_i,F_i;\mathbf{a}_i\rangle$ to obtain refined
approximations of a ground state. These do not possess a Gaussian
characteristic function but a lengthy yet straightforward
computation reveals that the corresponding characteristic function
$\chi_{|\psi\rangle\langle\psi|}(\bm{\xi})$ is a sum of Gaussian
functions with complex weights. Then it is immediately evident
that in this description we retain the ability of efficient
evaluation of all expectation values of polynomials in $X$ and
$P$. This allows one to establish an efficient algorithm for the
approximation of ground state properties of lattice Hamiltonians
that are polynomial in $X$ and $P$.

{\em Summary and Outlook: } We have introduced a new variational
method based on deformed weighted graph states to determine
approximations to ground states of strongly interacting spin
systems. The possibility to compute expectation values of local
observables efficiently, together with entanglement features
similar to those found in critical systems, make these states
promising candidates to approximate essential features of ground
states for systems with short range interactions in arbitrary
geometries and spatial dimensions. One can also generalize this
approach to describe the dynamics of such systems, systems with
long range interactions, disordered systems, dissipative systems,
systems at finite temperature and with infinite dimensional
constituents. In fact, generalizations of our method that deal
with these issues are possible and will be reported elsewhere.

We thank J.\,I.\ Cirac for valuable discussions
and J.\ Eisert for suggesting the use of the Anderson bound.
This work was supported by the FWF, the QIP-IRC funded by EPSRC
(GR/S82176/0), the European Union (QUPRODIS, OLAQUI, SCALA, QAP),
the DFG, the Leverhulme Trust (F/07 058/U), the Royal
Society, and the \"OAW through project APART (W.\,D.). Some
of the calculations have been carried using facilities of the University of Innsbruck's Konsortium
Hochleistungsrechnen.



\begin{thebibliography}{99}
\bibitem{footnoteDMRG}
We remark that for 1D systems, the accuracies appear to scale less well in the resources as for DMRG methods. However, our approach yields accurate results also for 2D and 3D systems.
\end{thebibliography}
\end{document}